# Evidence for metallic 1T phase, 3d$^1$ electronic configuration and charge density wave order in molecular-beam epitaxy grown monolayer VTe$_2$

*Ping Kwan Johnny Wong,[1,‡] Wen Zhang,[2,‡] Jun Zhou,[2] Fabio Bussolotti,[3] Xinmao Yin,[2] Lei Zhang,[1,2] Alpha T. N'Diaye,[4] Simon A. Morton,[4] Wei Chen,[1,2,5] Kuan Eng Johnson Goh,*[,2,3] Michel P. de Jong,[6] Yuan Ping Feng,*[,1,2] and Andrew Thye Shen Wee*[,1,2]*

[1]Centre for Advanced 2D Materials (CA2DM) and Graphene Research Centre (GRC), National University of Singapore, 6 Science Drive 2, Singapore 117546, Singapore

[2]Department of Physics, National University of Singapore, 2 Science Drive 3, Singapore 117542, Singapore

[3]Institute of Materials Research and Engineering (IMRE), Agency for Science, Technology and Research (A*Star), 2 Fusionopolis Way, Innovis, Singapore 138634, Singapore

[4]Advanced Light Source (ALS), Lawrence Berkeley National Laboratory, Berkeley CA94720, USA

[5]Department of Chemistry, National University of Singapore, 2 Science Drive 3, Singapore 117542, Singapore

[6]NanoElectronics Group, MESA+ Institute for Nanotechnology, P.O. Box 217, University of Twente, 7500 AE, Enschede, The Netherlands

ABSTRACT: **We present a combined experimental and theoretical study of monolayer $VTe_2$ grown on highly oriented pyrolytic graphite by molecular-beam epitaxy. Using various *in-situ* microscopic and spectroscopic techniques, including scanning tunneling microscopy/spectroscopy, synchrotron X-ray and angle-resolved photoemission, and X-ray absorption, together with theoretical analysis by density functional theory calculations, we demonstrate direct evidence of the metallic 1T phase and $3d^1$ electronic configuration in monolayer $VTe_2$ that also features a $(4 \times 4)$ charge density wave order at low temperatures. In contrast to previous theoretical predictions, our element-specific characterization by X-ray magnetic circular dichroism rules out a ferromagnetic order intrinsic to the monolayer. Our findings provide essential knowledge necessary for understanding this interesting yet less explored metallic monolayer in the emerging family of van der Waals magnets.**

INTRODUCTION

Two-dimensional transition-metal dichalcogenides (2D-TMDs)[1, 2] offer a promising avenue to addressing the imminent challenges brought by the continuous downscaling of conventional silicon-based electronic components. With over 40 different choices of compounds, the family of 2D-TMDs provides an extensive range of key electronic properties from metals, semiconductors to half-metals, and from magnets to superconductors.[3-5] These form the essential building blocks for emerging technologies that can leverage the unique assets of atomically thin materials – mechanical flexibility, chemical tunability, and coupled electronic degrees of freedom, such as charge, spin, orbital, valley, etc.[6-9]

Vanadium ditelluride, $VTe_2$, is an interesting yet rarely explored metallic 2D-TMD. Early work has reported an anomaly at 390–437 K in both heat capacity and resistivity measurements due to phase transition from a hexagonal 1T phase (at high temperature) to a monoclinic 1T' structure (at low-temperature).[10, 11] This 1T' structure denotes the distorted form of the 1T phase, with two-thirds of the V atoms in each monolayer displaced forming double zigzag chains with a $(3 \times 1)$ periodicity. This phase transition has been associated with charge density wave (CDW) formation, which renormalizes the telluride's electronic band via an enhanced overlap between the Te and V orbitals.[12] As a consequence, a d-electron count of V deviates from the initial $d^1$ configuration in the 1T phase, resulting in the 1T' phase.[13] Interestingly, the fact that $VTe_2$ exhibits a stronger interlayer coupling between its chalcogen atoms than its sulfide and selenide counterparts makes such phase transition behavior and its other properties more sensitive to layer thickness. Prior tight-binding band electronic structure calculations have indeed revealed considerable electron transfer between the Te p and V d bands in 1T-$VTe_2$,[13, 14] in which the Te $p_z$-orbitals plays a crucial role. As such, any disturbance of the Te $p_z$-orbitals or equivalently the Te-Te interlayer coupling would alter the $VTe_2$ properties. A relevant system that mimics such a scenario is $VTe_2$ intercalated with alkali-metals (Li and Na),[15] in which the 1T' phase was seen to occur at lower temperatures than its pristine form, with the values of structural parameters closer to that of the 1T structure.[15, 16] The lower phase transition temperature onset for intercalated $VTe_2$ underlines the profound effect of modified Te-Te interlayer coupling along the *c* crystal axis. Similar observations have been reported for multilayer $VTe_2$ nanosheets grown by chemical vapor deposition[17] and films by molecular-beam epitaxy (MBE),[18] respectively. In both cases, a 1T structure with bulk-like lattice constants is reported.

In this work, we demonstrate the MBE growth of monolayer $VTe_2$ on highly oriented pyrolytic graphite (HOPG), enabling direct access to the monolayer regime where the intrinsic Te-Te interlayer coupling is completely removed. Through a combination of *in-situ* microscopic and spectroscopic techniques, including scanning tunneling microscopy/spectroscopy (STM/STS), synchrotron-based photoemission spectroscopy (PES), angle-resolved PES (ARPES), and X-ray absorption spectroscopy (XAS), we give evidence of the metallic 1T phase and $d^1$ electronic configuration in monolayer $VTe_2$. Unlike the $(3 \times 1)$ double zigzag chain-like modulation in the bulk crystal, a $(4 \times 4)$ CDW order is observed, suggesting the significant effect of reduced dimensionality on the CDW instability in $VTe_2$. Density functional theory (DFT) calculation reveals the possible role of the graphitic substrate in this regard. We also use element-specific X-ray magnetic circular dichroism (XMCD) to address whether an intrinsic ferromagnetic order exists in monolayer $VTe_2$, as predicted by several theoretical studies.[19-21]

RESULTS AND DISCUSSION

**Film growth, structural and STM/STS measurements.** The 1T structure (space group: $P\bar{3}m1$) of VTe$_2$ is illustrated in **Figure 1**a, consisting of a plane of hexagonally arranged V atoms sandwiched by two atomic Te planes with in-plane and out-of-plane lattice parameters of 3.64 Å and 6.51 Å, respectively. These parameters have been confirmed by our STM measurements at 77 K for MBE-grown monolayer VTe$_2$ on HOPG (also see the atomic force microscopy image in **Figure S1** that shows the 2D growth of the monolayer). **Figure 1**b presents a large-scale STM image, indicating a step height of 8.5 Å for monolayer VTe$_2$ in **Figure 1**c, which is consistent with the combined thickness of the monolayer with a van der Waals gap on the substrate. The atomic-resolution STM image in **Figure 1**d captures a $(1 \times 1)$ hexagonal unit (marked in orange) with a lattice spacing of $3.6 \pm 0.1$ Å, which coexists with a $(4 \times 4)$ reconstructed pattern (marked in white) whose bias-dependence is shown in **Figure S2**. A Fourier transform of the STM image is shown in the inset (top-right corner) of **Figure 1**d, displaying alignment between the sharp $(1 \times 1)$ lattice peaks and the $(4 \times 4)$ spots. Our first STM observation of this reconstruction provides real-space identification that confirms a recently reported $(4 \times 4)$ pattern obtained from low-energy electron diffraction of monolayer VTe$_2$ on bilayer graphene.[22] According to ref. 22, this pattern is emerged below 186 K, due to CDW that opens an anisotropic energy-gap with a maximum size of 50 meV near the M point of the telluride's 2D Fermi surface; while, at the $\Gamma$ point, this gap is absent.[22] Our dI/dV spectrum in the inset (bottom-left corner) of **Figure 1**d displays consistent features with such an electronic structure.[22] Our spectrum, which represents the local density of states, reveals a differential conductance dip centered at the Fermi level, without a clearly resolved gap. The latter can be explained by the fact that the STS spectrum is momentum-integrated, with some of the momentum regions of monolayer VTe$_2$ gapped (along M–K) and others not ($\Gamma$ point). As measured at 77 K, we expect both spectral broadening and temperature-dependence of the CDW gap[22] to also affect the measured STS.

**Surface chemical properties probed by synchrotron-radiation PES. Figure 2** shows the core-level PES (a,b) and valence band (c) of as-grown monolayer $VTe_2$ (purple lines in the lower panel), and ambient-air exposed $VTe_2$ (green lines in the upper panel) for comparison. As shown in **Figure 2**a, the V 2p peaks are located at 512.9 and 520.5 eV in binding energy (BE), and the Te 3d peaks at 572.0 and 582.3 eV (**Figure 2**b). The positions of V 2p peaks suggest a valence state of +4 but are 0.3–0.5 eV lower than the case for monolayer $VSe_2$ on HOPG,[23] which we attribute to the lower electron affinity of Te (190.2 kJ/mol) than that of Se (195.0 kJ/mol). **Figure 2**c shows the valence band of monolayer $VTe_2$ obtained using a photon energy of 60 eV, indicating a metallic Fermi edge contributed mainly by V 3d states. The spectral features 1–6 eV below the Fermi edge are related to the Te 4p bands and the graphite substrate. We also extract, from the secondary electron cutoff, a work function value of 4.7 eV (**Figure 2**d), which is ~0.3 eV lower than that for monolayer $VSe_2$ prepared on HOPG under similar growth conditions.[23]

The different trends of the green lines (upper) from those of purple lines (lower) indicates that the intrinsic properties of monolayer $VTe_2$ are sensitive to ambient exposure. Notably, oxide formation leads to major changes in the monolayer surface chemical properties, as evidenced by the significant O 1s peak, shifting of the V 2p and Te 3d core-levels to higher BE, loss of spectral weight of the initial metallic Fermi edge, and reduced work function by ~0.2 eV. These spectral changes are in general similar to those observed for air-exposed monolayer $VSe_2$.[23]

**Electronic band structure and phase identification by ARPES. Figure 3**a,b shows the ARPES intensity maps of monolayer $VTe_2$ measured at 297 K and 11 K, respectively. Due to the small VTe$_2$ domain sizes (a few hundred nm; see **Figure S1**) relative to the ARPES detection area (~800 μm), these measured bands are averaged over different crystal domains and thus weak. Yet, near the Fermi level, one observes in both the maps a weakly dispersive V 3d band and a set of degenerate Te 4p bands with a strong downward dispersion. Overlays of the experimental bands at 11 K with those calculated by DFT suggest the monolayer's 1T structure (**Figure 3**c).[22] Such a consistency is on the other hand not achieved when compared to the calculated 2H dispersive bands (**Figure S3**). The fact that a single d-band rather than exchange-split bands is observed in our case also excludes a ferromagnetic ground state as consistently predicted by theory.[19-21] We shall return to this discussion with further evidence by XMCD measurements for the monolayer.

**Figure 3**d shows the effect of thermal broadening in the normalized energy distribution curves (EDCs) around the Γ point. Upon cooling to 11 K from 300 K, the peak width of the EDC is evidently reduced, yet without a leading-edge midpoint shift, indicating absence of a CDW-gap opening around the Γ point, consistent with the previous ARPES study[22] and our STS data in **Figure 1**d.

**Vanadium d-orbital electronic configuration and absence of intrinsic ferromagnetic order by XAS and XMCD. Figure 4**a shows splitting of the 3d degenerate orbitals of V ions in VTe$_2$ into two sets of triplet t$_{2g}$ and doublet e$_g$ states in the presence of an octahedral crystal field. In this d$^1$ odd system (V$^{4+}$), the unpaired electron preferentially fills the t$_{2g}$ states with relatively low orbital energy. We probed this configuration by means of V L$_{2,3}$-edge XAS in the total electron yield (TEY) mode (**Figure 4**b). The corresponding spectrum shown in the upper panel of **Figure 4**c consists of two main absorption peaks at 518 and 524 eV, corresponding to dipole-allowed transitions from the spin-orbit split V 2p$_{3/2}$ and 2p$_{1/2}$ core-levels to the 3d unoccupied states of monolayer VTe$_2$. The fine structures at energies below the main peaks are remnant of the atomic

multiplets. Being similar to those measured for bulk $VS_2$[24] and monolayer $VSe_2$,[23] these spectral features give strong evidence of the 1T phase and $d^1$ electronic configuration of monolayer $VTe_2$.

We also carried out V $L_{2,3}$-edge XMCD measurements at a temperature range of 16–300 K by measuring XAS and reversing the external magnetic field direction (± 1 Tesla). Contrary to predictions of ferromagnetic exchange splitting in the V d-bands by DFT calculations,[19-21] our monolayer possesses negligible XMCD contrast (lower panel of **Figure 4**c), thus suggesting a lack of intrinsic ferromagnetic order. We note that ferromagnetic signals have been reported for CVD-grown $VTe_2$ nanoplates on $SiO_2$/Si, characterized by vibrating sample magnetometry (VSM).[17] As a conventional magnetic measurement tool, VSM characterizes the macromagnetism of the whole sample, including the contribution from the substrate and probably from extrinsic perturbations. For instance, the substrate (Si, HOPG, etc.) usually contributes various magnetic signals.[25, 26] To subtract such extrinsic contributions is tricky, especially due to the atomically thin nature of the 2D material, and different methods of subtraction could yield diverse magnetic moments.[27, 28] In this regard, the XMCD is an advantageous tool, in that it is element-specific, which guarantees the observed magnetic contrast to be intrinsic to V. In particular, our XMCD results are supportive of our ARPES data in **Figure 3**, showing no evidence of exchange-split bands, which would otherwise exist in intrinsically ferromagnetic materials. We hypothesize that this is related to the CDW instability in monolayer $VTe_2$, which not only competes with but suppresses the ferromagnetic ground state predicted by DFT, as similarly reported for monolayer $VSe_2$.[29, 30]

**Discussion of the substrate-effects in suppressing the 1T' phase.** Fermi surface nesting and electron-phonon interaction have been invoked previously to account for the different CDW orders in monolayer $VTe_2$ and its bulk counterpart.[18, 22] Furthermore, our DFT calculations reveal possible substrate effects that could weaken the double zigzag chain modulation in the 1T' phase.

Specifically, we have considered four situations here: (1) freestanding 1T monolayer, (2) 1T monolayer relaxed on graphene, (3) freestanding 1T' monolayer, and (4) 1T' monolayers relaxed on graphene. We notice that no matter with the substrate or none, the 1T' phases (3 & 4; **Figure 5**a) are always more energetically stable than the 1T phases (1 & 2), which is however not observed experimentally by us. Quantitatively, the relative energy difference between these phases is 0.1 eV per super cell for the freestanding case, and ~0.08 eV per super cell for the on-graphene case. The small discrepancy of ~0.02 eV in the relative energy difference could be related to charge transfer from graphene to monolayer $VTe_2$, as shown in **Figure 5**b. Back to the abovementioned inconsistence between the calculation and our experimental observation, we continue to examine the substrate effect in suppressing the 1T' phase. As illustrated in **Figure 5**a, the V atoms in the 1T' monolayer initially displace along one lattice axis, the *b* axis in this work, forming the double zigzag strips. However, when relaxed on graphene, these regular displacements disappear and more random distortions are observed along both the *a* and *b* axes, as shown in **Figure 5**a. These considerably reduce both the Te-Te height profile (**Figure 5**c) and the V-strips separation (**Figure 5**d), and in turn the 1T' structural modulations. We speculate that, when a multilayer graphene is used as a substrate, the structural modulations will be further smeared out, which may probably explain why the 1T' phase is not observed in our monolayer on HOPG.

CONCLUSIONS

In summary, we present direct microscopic and spectroscopic evidence of metallic 1T phase and $d^1$ electronic configuration in monolayer $VTe_2$. Unlike the double zigzag chained structure observed in the bulk, we obtain a $(4 \times 4)$ CDW reconstruction pattern. DFT calculations reveal the possible role of the graphitic substrate in suppressing the double zigzag structure. Regarding the magnetism of monolayer $VTe_2$, our XMCD data excludes an intrinsic ferromagnetic ordering, in disagreement with previous theoretical predictions. Our findings provide new knowledge of this metallic vanadium dichalcogenide monolayer, potentially interesting for the explorations of exotic quantum phenomena and device applications related to phase transitions.

METHODS

**Molecular-beam epitaxy of monolayer VTe$_2$.** Monolayer VTe$_2$ films were grown on HOPG in a custom-built MBE chamber with a base pressure of better than $1 \times 10^{-9}$ mbar. The substrates were prepared by *in-situ* cleavage followed by annealing at 820 K for at least 120 min. High-purity V and Te were evaporated from an electron-beam evaporator and a standard Knudsen cell, respectively. The Te/V flux ratio was controlled to be >10. During the growth process, the substrate temperature was kept at 650 K. To protect the samples against ambient contaminations during *ex-situ* transport to other ultrahigh vacuum (UHV) measurement systems, a Se/Te bilayer was deposited on the samples as a cap. For subsequent characterization by PES, ARPES, and XAS/XMCD, the cap was desorbed in UHV at 500 K. It is noteworthy that a Se 3d core-level peak was evident by PES after thermal desorption of a pure Se cap (**Figure S4**), suggesting possible Se contamination in the monolayer VTe$_2$ film, as such signal was quite robust and remained even at relatively high temperatures.

**Scanning tunneling microscopy and spectroscopy.** STM measurements were carried out in a custom-built multi-chamber UHV system housing an Omicron low temperature-STM interfaced with a Nanonis controller. The base pressure was better than $10^{-10}$ mbar. A chemically etched tungsten tip was used, and the sample was kept at 77 K during all the measurements. STM images were recorded in constant current mode. For dI/dV spectra, the tunneling current was obtained using the lock-in technique. Note that the bias voltage is applied on the STM tip; hence, negative values correspond to conduction bands and positive values correspond to valence bands.

Each curve was obtained by averaging hundreds of individual spectra acquired at random locations of a specific STM image.

**Synchrotron-radiation photoemission.** PES measurements were performed at 300 K at the SINS beamline of the Singapore Synchrotron Light Source (SSLS), which covers the photon energy range from 50 to 1200 eV. A Scienta SES-200 spectrometer was used to collect the spectroscopic data at normal emission, while the X-ray beam was set with an incident angle of 45° relative to the sample surface. A bias voltage of −7.0 eV was applied to the sample during work function measurement in order to negate the effect of the analyzer work function. The binding energy of the data was calibrated using the 4f core-levels and Fermi edge of a reference Au foil.

**Angle-resolved photoemission.** The ARPES measurements were collected with HeI$\alpha$ (h$\nu$ = 21.218 eV) radiation source (SCIENTA VUV5k). The photoelectrons were analyzed in the plane of incidence with a high energy and angular resolution SCIENTA DA30L analyzer. The angular detection range spans ±15° with respect to the spectrometer lens axes. Wider angular limits were obtained by rotating the sample with respect to the analyzer lens entrance axes. During ARPES acquisition the total energy resolution was set to 20 meV, with the angular resolution being better than 0.2°. The binding energy scale was referred to the Fermi level ($E_F$) as measured for a clean gold substrate.

**X-ray absorption spectroscopy and magnetic circular dichroism.** XAS and XMCD measurements were carried out at the beamline 6.3.1 of the Advanced Light Source (ALS). The spectra were collected with sample temperatures ranging from 16 to 300 K in TEY mode, in which the sample drain current was recorded as a function of the photon energy. The angle of incidence of the photon beam was set to 45° relative to the sample surface. XMCD spectra were recorded with a fixed circular polarization of the X-rays and opposite magnetic fields up to ±1 T.

**Density-functional theory calculations.** The DFT calculations were performed using the VASP package, utilizing the projector augmented phase wave (PAW) method,[31] and the Perdew Burke and Ernzerhof (PBE) exchange-correlation functional.[32] To better estimate the interlayer dispersion interactions in the interface structure, the dispersion-corrected vdW-optB88 exchange-correlation functional was applied.[33-36] A separation of 20 Å between $VTe_2$ layers was found to be sufficient to represent an isolated monolayer. We employed a kinetic energy cutoff of 500 eV and a $\Gamma$-centered $21 \times 21 \times 1$ $k$-point mesh. The lattice parameters and the atomic positions were optimized until the forces on the atoms were less than 1 meV/Å. The relaxed lattice parameter is 3.61 Å, in good agreement with the experimental value. For the interfaces, we use a $3 \times 9$, $2 \times 6$, and $2 \times 2$ supercell for graphene, T phase and T' phase $VTe_2$, respectively, in which the strain is less than 2%. The graphene layer is kept fixed during the structure relaxation.

FIGURES.

**Figure 1. Lattice structure and STM/STS data of monolayer VTe₂ on HOPG.** (a) Lattice structure of 1T-VTe₂. (b) Large-scale STM image of the monolayer measured at 77 K (150 × 150 nm²; tip bias = −0.89 V, tunneling current = 68 pA). Marked by yellow arrows are residual cap. (c) The line profile shown in (b) with a monolayer step height of ∼8.5 Å. (d) Atomic-resolution STM image (10 × 10 nm²; tip bias = +0.1 V, tunneling current = 150 pA), which reveals an in-plane lattice parameter of 3.6 ± 0.1 Å, consistent with that in (a) for 1T-VTe₂. The additional (4 × 4) superstructure is observed in the STM image as well as in its FFT. The averaged STS curve (set-point: tip bias = +0.3 V, tunneling current = 60 pA, 625 Hz, 50 mV) of the monolayer shows no sign of a gap feature related to a CDW order.

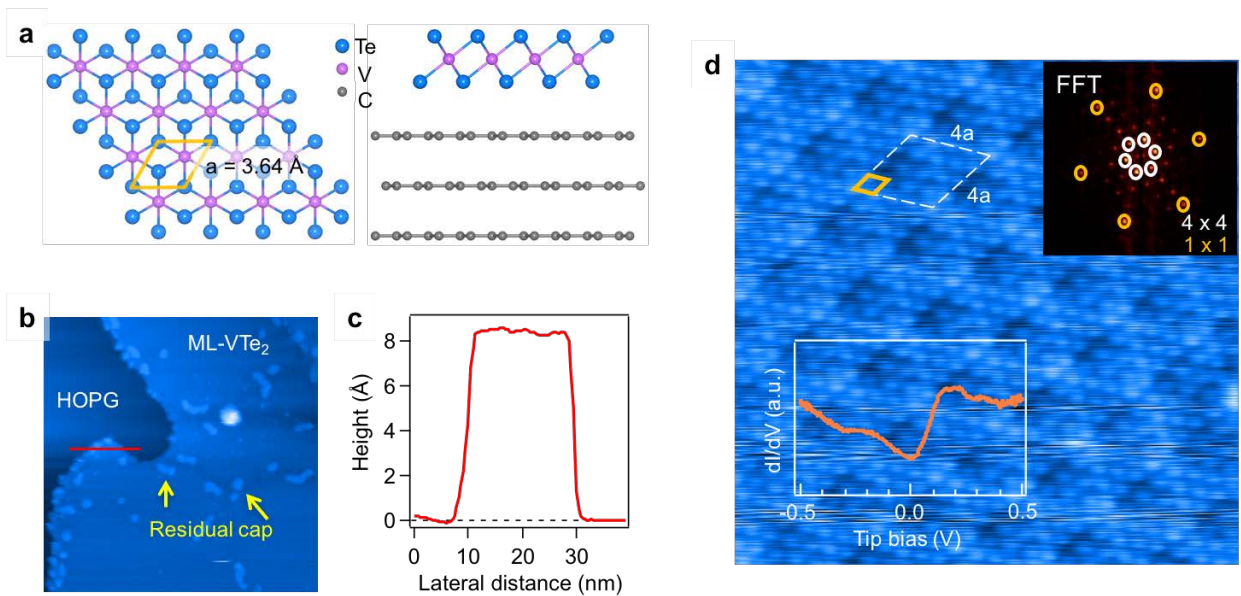

**Figure 2. PES data of monolayer VTe₂.** (a,b) V 2p and Te 3d core-levels of monolayer VTe₂. The V 2p$_{3/2}$ and 2p$_{1/2}$ peaks are seen to position at 512.9 eV and 520.5 eV, respectively, and the Te 3d$_{5/2}$ and 3d$_{3/2}$ peaks are at 572.0 eV and 582.3 eV, respectively. (c) The valence band of the monolayer shows a metallic Fermi edge contributed mainly by the V 3d states. The spectral features at 1−5 eV are derived from the Te 4p bands, while the broad peak beyond 6 eV is originated from the graphite substrate. (d) Work function extracted from the secondary electron cutoff of the monolayer, using a photon energy of 60 eV, is ~4.7 eV. The upper panel (data in green) shows the impacts of ambient-air-exposure. A peak shift to higher BE is evidenced for both the V 2p and Te 3d core-levels and the initial metallic Fermi edge loses most of its spectral weight. These changes are also accompanied by a work function decrease by 0.2 eV.

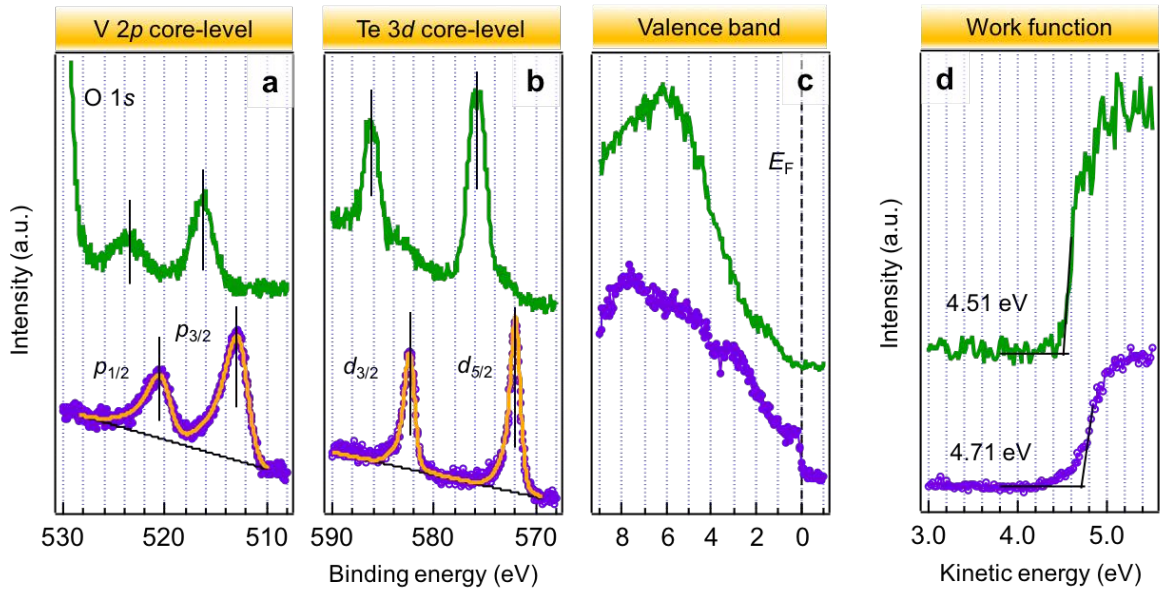

**Figure 3. ARPES results of monolayer VTe₂.** (a,b) Intensity maps measured at 300 K and 11 K, respectively, as a function of the surface momentum component $k_{//}$. (c) Overlay between the experimental (11 K) and calculated band dispersions. Note that the dispersions along the $\Gamma-M$ direction in the hexagonal Brillouin zone are in solid lines and those along the $\Gamma-K$ direction in dotted lines. Comparison between these bands confirms the monolayer's 1T structure. The observation of a single d-band also excludes the theoretically predicted ferromagnetic ground state. (d) Upon cooling to 11 K from 300 K, the EDC around the $\Gamma$ point, normalized by the FD function, reveals a smaller spectral width due to reduced thermal broadening. However, no leading-edge midpoint shift is observed in the EDCs, indicating no CDW gap opening at the $\Gamma$ point. For all measurements, zero binding energy represents the Fermi level position.

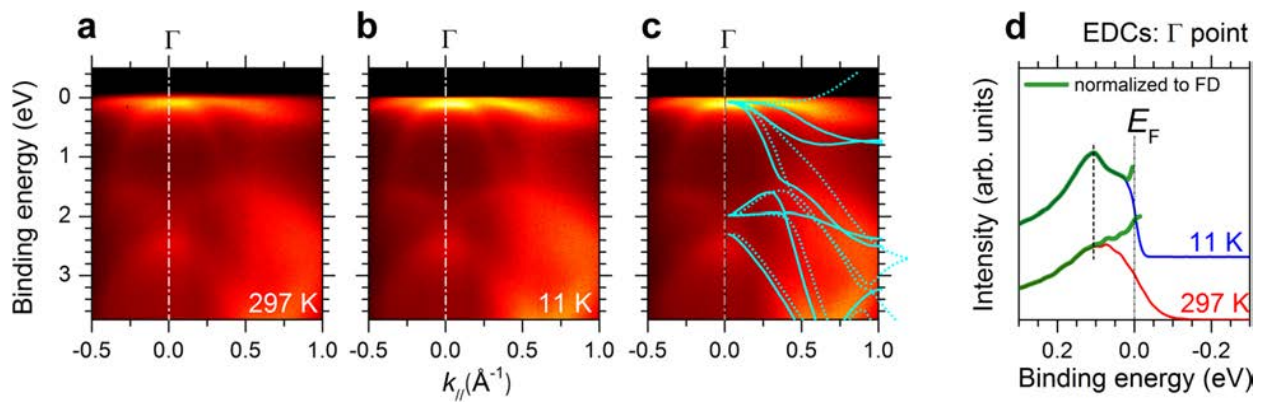

**Figure 4. V d$^1$ electronic configuration of 1T-VTe$_2$ and element-specific XAS/XMCD characterization of monolayer VTe$_2$.** (a) In an octahedral crystal-field, the 3d-orbitals of the V$^{4+}$ ion in 1T-VTe$_2$ are split into two sets of orbitals, e$_g$ and t$_{2g}$, separated by an energy of 10 D$q$. (b) TEY detection of the XAS/XMCD data illustrated in (c). (c) Upper panel shows the XAS spectra of the monolayer measured at 16 K. Spectra highlighted in solid and dotted lines are the XAS acquired with opposite external magnetic fields (±1 T), respectively. Distribution of the atomic multiplets observed for the monolayer provides strong evidence of the d$^1$ configuration depicted in (a). Lower panel in (c) shows the corresponding XMCD signal, which, within experimental error, suggests a lack of ferromagnetic coupling in monolayer VTe$_2$.

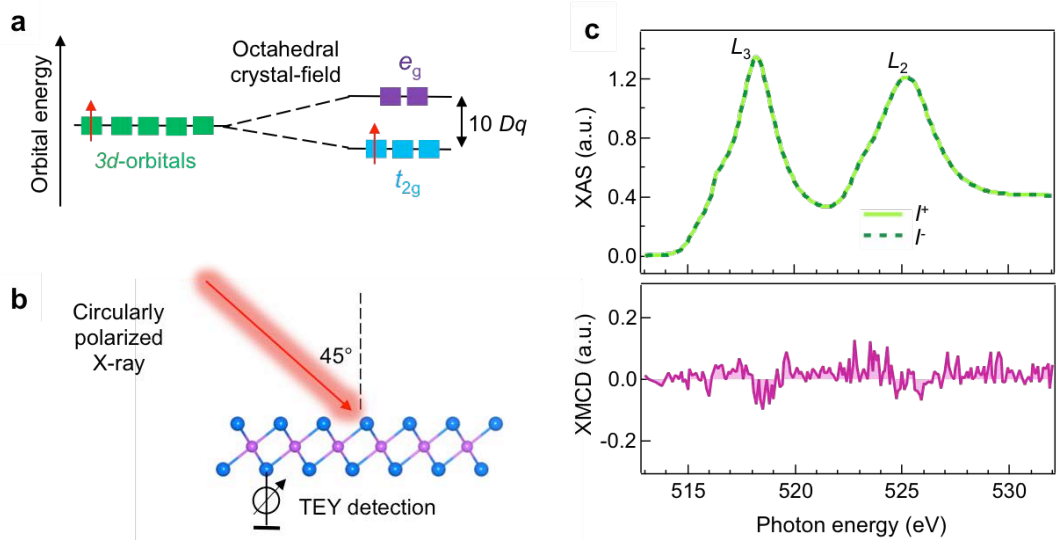

**Figure 5. Comparison of the DFT results for 1T' monolayer VTe₂ in free-standing form and on graphene.** (a) Structural guidance for free-standing T'-VTe₂ in side view. (b) Structural guidance for T'-VTe₂ on graphene with charge difference in isosurface value of $2.5 \times 10^{-4}$ e/Bohr³ in side view. The charge density difference is calculated by subtracting the sum of the charge densities of the VTe₂ and graphene layers from the total charge density of the whole interface slab. The blue and red area represents charge depletion and accumulation, respectively. (c) The height difference of Te atoms in the top sublayer of VTe₂ for free-standing and on-substrate form. The height of the lowest Te is taken as reference and is set as 0. The numbers in the horizontal axis correspond to the Te atoms shown in (a). For the case with graphene substrate, the Te atoms have different heights along $a$ axis and the average value is used. (d) The distance between the V strips along $a$ axis of VTe₂ for free-standing and on-substrate form. The numbers in the horizontal axis correspond to the V atoms shown in (a). For the case with graphene substrate, the V atoms have different coordinate along $a$ axis and the average value is used.

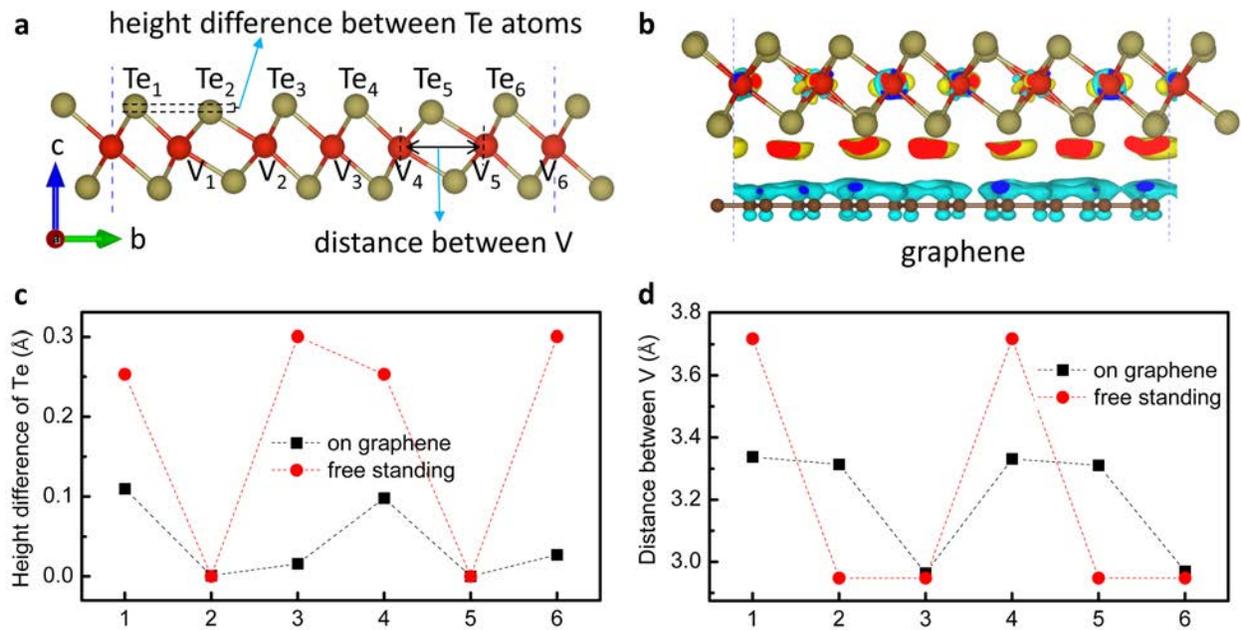

## ASSOCIATED CONTENT

**Supporting Information**. The supporting Information is available free of charge on the ACS

Publication website.

    *Ex-situ* atomic force microscopy image; Bias-dependent STM images of monolayer $VTe_2$; Overlay of experimental ARPES bands with theoretical bands of $2H\text{-}VTe_2$; Se 3d XPS peak with pure Se as a cap (PDF)

    The authors declare no competing financial interests.

## AUTHOR INFORMATION

**Corresponding Author**

*E-mail: gohj@imre.a-star.edu.sg

*E-mail: phyfyp@nus.edu.sg

*E-mail: phyweets@nus.edu.sg

**Author Contributions**

‡These authors contributed equally.

## ACKNOWLEDGMENT

    We acknowledge financial support from the Singapore Ministry of Education Tier 2 grants (MOE2016-T2-2-110 and R-143-000-652-112), the National Research Foundation Medium Sized Centre Programme (R-723-000-001-112) and the A*STAR 2D PHAROS Grant (1527000016 and R-144-000-359-305). This research used resources of the Advanced Light Source, which is a DOE Office of Science User Facility under contract no. DE-AC02-05CH11231. The authors also acknowledge the Singapore Synchrotron Light Source (SSLS), a National Research Infrastructure